
%
%
\magnification=1200
\voffset=0 true mm
\hoffset=0 true in
\hsize=6.5 true in
\vsize=8.5 true in
\normalbaselineskip=13pt
\def\doublespace{\baselineskip=20pt plus 3pt\message{double space}}
\def\singlespace{\baselineskip=13pt\message{single space}}
\let\spacing=\singlespace
\parindent=1.0 true cm



\newcount\equationumber \newcount\sectionumber 
\sectionumber=1 \equationumber=1               
\def\setsection{\global\advance\sectionumber by1 \equationumber=1} 
\def\numbe{{{\number\sectionumber}{.}\number\equationumber}
                            \global\advance\equationumber by1}
\def\numberit{\eqno{(\number\equationumber)} \global\advance\equationumber by1}
%
\def\numberal{(\number\equationumber)\global\advance\equationumber by1}
%
%
%
%


\def\ccf#1{\,\vcenter{\normalbaselines
    \ialign{\hfil$##$\hfil&&$\>\hfil ##$\hfil\crcr
      \mathstrut\crcr\noalign{\kern-\baselineskip}
      #1\crcr\mathstrut\crcr\noalign{\kern-\baselineskip}}}\,}
\def\scf#1{\,\vcenter{\baselineskip=9pt
    \ialign{\hfil$##$\hfil&&$\>\hfil ##$\hfil\crcr
      \vphantom(\crcr\noalign{\kern-\baselineskip}
      #1\crcr\mathstrut\crcr\noalign{\kern-\baselineskip}}}\,}

\def\small3j#1#2#3#4#5#6{\def\st{\scriptstyle} 
   \bigl(\scf{\st#1&\st#2&\st#3\cr
           \st#4&\st#5&\st#6\cr} \bigr)}


\def\ref#1{$^{#1)}$}    


\def\upa#1{\raise 1pt\hbox{\sevenrm #1}}
\def\dna#1{\lower 1pt\hbox{\sevenrm #1}}
\def\dnb#1{\lower 2pt\hbox{$\scriptstyle #1$}}
\def\dnc#1{\lower 3pt\hbox{$\scriptstyle #1$}}
\def\upb#1{\raise 2pt\hbox{$\scriptstyle #1$}}
\def\upc#1{\raise 3pt\hbox{$\scriptstyle #1$}}
\def\hprime{\raise 2pt\hbox{$\scriptstyle \prime$}}
\def\ccom{\,\raise2pt\hbox{,}}


\def\asymptotically#1{\;\rlap{\lower 4pt\hbox to 2.0 true cm{
    \hfil\sevenrm  #1 \hfil}}
   \hbox{$\relbar\joinrel\relbar\joinrel\relbar\joinrel
     \relbar\joinrel\relbar\joinrel\longrightarrow\;$}}
\def\Asymptotically#1{\;\rlap{\lower 4pt\hbox to 3.0 true cm{
    \hfil\sevenrm  #1 \hfil}}
   \hbox{$\relbar\joinrel\relbar\joinrel\relbar\joinrel\relbar\joinrel
     \relbar\joinrel\relbar\joinrel\relbar\joinrel\relbar\joinrel
     \relbar\joinrel\relbar\joinrel\longrightarrow$\;}}

\catcode`@=11
\def\C@ncel#1#2{\ooalign{$\hfil#1\mkern2mu/\hfil$\crcr$#1#2$}}
\def\gf#1{\mathrel{\mathpalette\c@ncel#1}}      
\def\Gf#1{\mathrel{\mathpalette\C@ncel#1}}      

\def\gapx{\lower 2pt \hbox{$\buildrel>\over{\scriptstyle{\sim}}$}}
\def\lapx{\lower 2pt \hbox{$\buildrel<\over{\scriptstyle{\sim}}$}}

\def\nablaleft{\hbox{\raise 6pt\rlap{{\kern-1pt$\leftarrow$}}{$\nabla$}}}
\def\nablaright{\hbox{\raise 6pt\rlap{{\kern-1pt$\rightarrow$}}{$\nabla$}}}
\def\nablaboth{\hbox{\raise 6pt\rlap{{\kern-1pt$\leftrightarrow$}}{$\nabla$}}}

\def\boks#1#2{{\hsize=#1 true cm\parindent=0pt   
  {\vbox{\hrule height1pt \hbox
    {\vrule width1pt \kern3pt\raise 3pt\vbox{\kern3pt{#2}}\kern3pt
    \vrule width1pt}\hrule height1pt}}}}

\def\heading{ }
\def\range{ }

\def\body{\vfill\eject\parindent=1.0 true cm
 \ifx\spacing\singlespace\singlespace\else\doublespace\fi}
\def\title#1{\centerline{{\bf #1}}}

\def\today{\ifcase\month\or
  January\or February\or March\or April\or May\or June\or
  July\or August\or September\or October\or November\or December\fi
  \space\number\day, \number\year}
\let\date=\today
\newcount\hour \newcount\minute
\countdef\hour=56
\countdef\minute=57
\hour=\time
  \divide\hour by 60
  \minute=\time
  \count58=\hour
  \multiply\count58 by 60
  \advance\minute by -\count58

\def\sectionskip{\penalty-500\vskip24pt plus12pt minus6pt}

\def\sec{\hbox{\lower 1pt\rlap{{\sixrm S}}{\hbox{\raise 1pt\hbox{\sixrm S}}}}}
\def\section#1\par{\goodbreak\message{#1}
    \sectionskip\nobreak\noindent{\bf #1}\vskip0.3cm \noindent}

\nopagenumbers
\headline={\ifnum\pageno=\count31\frontheadline
  \else{\ifnum\pageno=0\frontheadline
     \else{{\raise 2pt\hbox to \hsize{\paperhead}}}\fi}\fi}

\footline={\centerline{\sevenbf \folio}}

\def\frontheadline{\sevenbf \hfil}
\def\paperhead{\sevenbf \heading\ \range\hfil\folio}
\newdimen\pagewidth \newdimen\pageheight \newdimen\ruleht
\maxdepth=2.2pt
\pagewidth=\hsize \pageheight=\vsize \ruleht=.5pt

\def\onepageout#1{\shipout\vbox{ 
    \offinterlineskip 
  \makeheadline
    \vbox to \pageheight{
         #1 
 \ifnum\pageno=\count31{\vskip 21pt\line{\the\footline}}\fi
 \ifvoid\footins\else 
 \vskip\skip\footins \kern-3pt
 \hrule height\ruleht width\pagewidth \kern-\ruleht \kern3pt
 \unvbox\footins\fi
 \boxmaxdepth=\maxdepth}
 \advancepageno}}

\output{\onepageout{\pagecontents}}

\count31=-1
\topskip 0.7 true cm

\centerline{\bf Do Black Holes Exist?}
\centerline{\bf J. W. Moffat}
\centerline{\bf Department of Physics}
\centerline{\bf University of Toronto}
\centerline{\bf Toronto, Ontario M5S 1A7}
\centerline{\bf Canada}
\vskip 1 true in
\centerline{\bf February, 1993}
\vskip 3 true in
{\bf UTPT-93-04}
\vskip 0.2 true in
{\bf e-mail: Moffat@medb.physics.utoronto.ca}
\par\vfil\eject
\centerline{\bf Do Black Holes Exist?}
\centerline{\bf J. W. Moffat}
\centerline{\bf Department of Physics}
\centerline{\bf University of Toronto}
\centerline{\bf Toronto, Ontario M5S 1A7}
\centerline{\bf Canada}
\vskip 0.7 true in
\centerline{\bf Abstract}
\vskip 0.2 true in
The problem of information loss in black hole formation and the associated
violations of basic laws of physics, such as conservation of energy,
causality and unitarity, are avoided in the nonsymmetric gravitational theory,
if the NGT charge of a black hole and its mass satisfy an inequality that does
not violate any known experimental data and allows the existence of white
dwarfs and neutron stars.
\par\vfil\eject

The occurrence of a naked singularity in classical general relativity (GR)
would lead to a breakdown of predictability of the theory at the singularity,
and a general breakdown of the classical laws of physics. There are
indications that GR admits naked singularity solutions$^{1}$, causing an unease
about the possibility of a breakdown of the cosmic censorship hypothesis$^{2}$.

Hawking's discovery of black hole radiance,$^{3}$ led him to
state that the accepted principles of relativity and quantum
mechanics cannot be reconciled with this phenomenon$^{4}$. All suggested
resolutions of the paradox result
in violations of cherished physical laws, such as the conservation of energy,
causality and unitarity$^{5}$. Consider a pure quantum state describing the
infalling matter of a star. Such a state can be described by a density matrix,
$\rho=\vert \psi><\psi\vert$ with vanishing entropy, $S=-\hbox{Tr}\rho
\hbox{ln}\rho$.
If, for sufficiently large $M$, the falling matter collapses to form a
black hole, it will begin emitting Hawking radiation. If the back-reaction
of the emitted radiation on the geometry is neglected, an approximation
expected
to hold until the mass $M$ becomes comparable to the Planck mass $M_{Pl}$,
then the radiation is thermal and is described by a mixed quantum state.
The radiation is approximately described by a thermal density matrix with a
nonzero entropy, $S\sim M^2/M_{Pl}^2$. The entropy is interpreted as an
absence of information, and because it is nonvanishing, the information
contained in the initial quantum state that fell into the black hole does
not subsequently escape in the form of Hawking radiation. Information has been
lost and unitarity has been violated. If the black hole evaporates completely,
then an initially pure quantum state that was precisely known becomes a mixed
state, and we can only assign probabilities to what the final state will be.
Thus, information is destroyed in principle.

The problem begins with the formation of an arbitrarily large, classical
black hole, which emits Hawking radiation. We cannot expect to invoke arguments
based on some, as yet, unknown quantum gravity theory to cure the problem,
since
such a theory is only operative at the Planck length, $L_{Pl}\sim 10^{-33}$ cm.
This would suggest that we must seek an alternative to Einstein's
gravitational theory i.e. we must modify the short distance behavior
of gravity and prevent the formation of black holes. If black holes cannot
form, then there is no Hawking radiation or information loss and the paradox
has been removed. In order to prevent singularities and black holes from
existing, we must make gravity repulsive in certain situations. The simplest
modification of Einstein's theory is the Brans-Dicke scalar-tensor
theory$^{6}$.
However, in this theory gravity is always attractive, so that the theory
predicts singularities and black holes as in GR.

The nonsymmetric gravitational theory (NGT) is a mathematically and physically
consistent theory of gravitation, which contains GR in a well-defined limit
$^{7}$. In this theory, the $g_{\mu\nu}$ decomposes according to the rule:
$$
g_{\mu\nu} = g_{(\mu\nu)} + g_{[\mu\nu]},
\numberit
$$
where $g_{(\mu\nu)} = {1\over 2} (g_{\mu\nu} + g_{\nu\mu})$ and $g_{[\mu\nu]} =
{1\over  2} (g_{\mu\nu} - g_{\nu\mu})$. The affine connection
$W^{\lambda}_{\mu\nu}$ decomposes according to
$$
W^\lambda_{\mu\nu} = W^\lambda_{(\mu\nu)} +
W^\lambda_{[\mu\nu]}.
\numberit
$$
The NGT contracted curvature tensor in terms of the $W$-connection is given by
$$
R_{\mu\nu}(W)=W^\beta_{\mu\nu,\beta} - {1\over
2}(W^\beta_{\mu\beta,\nu}+W^\beta_{\nu\beta,\mu}) -
W^\beta_{\alpha\nu}W^\alpha_{\mu\beta} +
W^\beta_{\alpha\beta}W^\alpha_{\mu\nu}.
\numberit
$$

The field equations are
$$
G_{\mu\nu} (W) = 8\pi G T_{\mu\nu},
\numberit
$$
$$
{{\bf g}^{[\mu\nu]}}_{,\nu} = 4\pi {\bf S}^\mu,
\numberit
$$
where $G_{\mu\nu} = R_{\mu\nu} - {1\over 2} g_{\mu\nu} R$.

The energy-momentum tensor for a perfect fluid in NGT is of the form$^{8}$:
$$
T^{\mu\nu} = (\rho+p)u^\mu u^\nu - p g^{\mu\nu},
\numberit
$$
where $\rho$ and $p$ are the density and pressure and $u^\mu = dx^\mu/ds$.
We interpret $S^\mu$ as the conserved particle number of the fluid:
$$
S^\mu = \sum_i f_i^2 n_i u^\mu,
\numberit
$$
where $f_i^2$ are the coupling constants associated with stable fermion
particles $i$, and $n_i$ denotes the particle density. The net NGT charge
associated
with a body is defined by
$$
\ell^2 =\int {\bf S}^0 d^3x
\numberit
$$
and because of the identity:
$$
{{\bf S}^\nu}_{,\nu}=0,
\numberit
$$
the $\ell^2$ of a body is a conserved quantity. The matter response
equations are
$$
{1\over 2} \left(g_{\sigma\rho}{\bf T}^{\sigma\alpha} + g_{\rho\sigma}{\bf
T}^{\alpha\sigma}\right)_{,\alpha} - {1\over 2} g_{\alpha\beta,\rho}{\bf
T}^{\alpha\beta} + {1\over 3} W_{[\rho,\nu]}{\bf S}^\nu = 0,
\numberit
$$
where $W_\mu=W^\alpha_{[\mu\alpha]}$.

A cooling star of mass greater than a few solar masses cannot reach
equilibrium as either a white dwarf or a neutron star in GR, and
a similar situation prevails in NGT$^{9}$.
Savaria has found solutions to the time dependent, spherically symmetric
collapse equations in NGT$^{10}$.
He proved that, if the final state of the collapsing matter is described
by an equation
of state for which the repulsive NGT force dominates, then it is possible to
produce a bounce mechanism which can prevent the formation of a black hole,
without violating the Hawking-Penrose$^{11}$ positivity condition:
$T_{\mu\nu}u^\mu u^\nu > 0$.

In NGT, there are no negative energy contributions, even in the presence
of repulsive forces. This is an important aspect of the theory, which
makes it a viable modification of GR. The flux of gravitational waves at
infinity is the same as in GR, i.e. it is finite and positive definite$^{12}$.
For the case of a time dependent, spherically symmetric spacetime, the
line element in polar coordinates is given by
$$
ds^2=\gamma(r,t)dt^2-\alpha(r,t)dr^2-R^2(r,t)d\Omega^2,
\numberit
$$
where $d\Omega^2=d\theta^2+\hbox{sin}^2\theta d\phi^2$. We choose $w(r,t)=
g_{[10]}\not=0$ and $f(r,t)=g_{[23]}=0$ in order to satisfy the physical
boundary condition of asymptotic flatness at infinity.

 From (10), we get for a perfect fluid the conservation law:
$$
\dot \rho= - (\rho+p){\partial\over \partial t}\hbox{ln}(\gamma^{-1/2}
\sqrt{-g}),
\numberit
$$
and
$$
p^{\prime} = {1\over 2}(\rho+p){\gamma^{\prime}\over \gamma}
-{1\over 6\pi}{s\over wR}W_{[0,1]},
\numberit
$$
where $\dot \rho= \partial \rho/\partial t$ and $p^{\prime}=
\partial p/\partial r$. Moreover,
$$
s(r,t)={2\pi RwS^0\over R^{\prime}}={d\ell^4\over dR^4}.
\numberit
$$

For the spherically symmetric collapse of a pressureless fluid, the
solutions for $\alpha(r,t)$ and $w(r,t)$ are given by
$$
\alpha(r,t)={[R^{\prime}(1+\ell^4/R^4)]^2\over (1-{2GM\over r})(1
+{\ell^4\over r^4})},
\numberit
$$
$$
w(r,t)={\ell^2\over R^2}R^\prime\biggl[(1-{2GM\over r})(1+{\ell^4\over r^4})
\biggr]^{-1/2}.
\numberit
$$
The collapse cannot be isotropic unless $\ell^2=0$ (the GR limit).
It can be shown that for a pressureless fluid,
$dM/ds=0$, so that $M=M(r)$ is a constant of the motion as in GR.  We
identify $M(r)$ as the mass contained within a sphere
of radius $r$. Moreover, due to (9), $\ell^2 = \ell^2(r)$ is also a
constant of the motion and will be identified as the NGT charge contained
within the sphere. Another constant of the motion is given by
$$
E(r)=\biggl[\biggl(1+ U^2 - {2GM\over R}\biggr)
\biggl(1+{\ell^4\over R^4}\biggr)\biggr]^{1/2} -
{1\over 2}{M\ell^2\ell_t^2\over m_tR^4},
\numberit
$$
where $U=dR/ds$ and $\ell^2_t$ and $m_t$ are the $\ell^2$ charge and
mass of a test particle.

We are interested in the case when the repulsive NGT forces are at a
maximum, which corresponds to neglecting the tensor force contribution
in (10) and (17). We assume that the fluid is at rest
at $t=0$:
$$
R(r,0)=r,\quad\dot R(r,0) = 0.
\numberit
$$
Then, (17) yields
$$
E(r)=[(1-{2GM\over r})(1+{\ell^4\over r^4})]^{1/2}.
\numberit
$$
Let us further assume that the gravitational field is weak, so that
$$
{2GM\over r} << 1,\quad {\ell^4\over r^4} << 1.
\numberit
$$
Solving the equation:
$$
\dot R(r,t)=0,
\numberit
$$
we find that the collapsing star with initial density $\rho(r,0)$ and zero
pressure will not form an event horizon and a black hole when:
$$
\ell \geq \biggl({8\over 3^{3/4}}\biggr)GM.
\numberit
$$
When $M=M_{\odot}$, the collapse is halted before the event horizon is reached
for $\ell \geq 5.2$ km. This result agrees with the simplified
case of the radial collapse of a shell of test particles in the exterior
static,
spherically symmetric NGT metric$^{7}$.

It is not expected that the basic mechanism preventing black hole formation
in stellar collapse will be significantly modified when the approximations
made above
are relaxed, but further work must be carried out to investigate this issue.

For a Wheeler GEON (gravitational electromagnetic entity)$^{13}$, it was shown
by Wheeler$^{14}$, and Brill and Hartle$^{15}$, that such objects are
unstable i.e. they will either disperse or collapse. Although it is unlikely
that such objects exist, we must consider their behaviour in NGT.
Because the photon number density $n_\gamma$ is not conserved,
we have $\ell^2_\gamma=0$ and $g_{[\mu\nu]}=0$, and NGT reduces to GR for
photon GEONS. Therefore, if photon GEONS exist, we cannot prevent their
collapse to a black hole with the resultant loss of information due to Hawking
radiation. On the other
hand, for neutrino GEONS$^{16}$, NGT will prevent their collapse to a black
hole, because the NGT charge for neutrinos, $\ell^2_\nu \not=0$, thereby
guaranteeing that these
objects can be prevented from collapsing to black holes. However, as with
photon
GEONS, we consider that such systems are highly unlikely to exist as
astrophysical objects.

We shall postulate as a general principle in Nature, that since the
initial big bang at t=0, matter always dominated over anti-matter. Thus,
matter genesis was initiated at t=0 by means of a violation of particle number
conservation and an associated CP violation. In NGT, this breaks the principle
of equivalence, since the excess of matter coupled to
the NGT field $g_{[\mu\nu]}$, violates the equivalence principle.  We also
postulate that at the end-point of gravitational collapse, the repulsive
NGT force dominates, so as to prevent the formation of black holes of
arbitrarily large or small mass. A consequence of this is that no naked,
local singularities of the type generated by gravitational collapse can
ever be observed in Nature, nor can event horizons be produced together with
Hawking radiation. In NGT there will be no absolute loss of information,
as occurs in GR, and the laws of quantum mechanics and thermodynamics will
not be violated.

The values of the NGT charge necessary to prevent the formation of
black holes are not expected to contradict any experiments
in the solar system$^{7}$ (this is certainly true for pressureless collapse),
nor will they prevent the formation of white dwarfs or neutron stars$^{9}$.
It will be interesting to
investigate the astrophysical consequences of the objects that form when
stellar collapse is halted in NGT, and to ascertain whether stable systems
can be formed with possibly unique observational signatures that could
distinguish them from black holes.
\vskip 0.3 true in
{\bf Acknowledgements}
\vskip 0.2 true in
This work was supported by the Natural Sciences and Engineering Research
Council of Canada. I thank M. Clayton, N. Cornish, P. Savaria and D. Tatarski
for helpful and stimulating discussions.
\vskip 0.5 true in
\centerline{\bf References}
\vskip 0.3 true in
\item{1.}{For some references on this subject, see: D. Eardly and
L. Smarr, Phys. Rev. D{\bf 19}, 2239 (1979);
D. Christodoulou, Commun. Math. Phys. {\bf 93}, 171 (1984);
R. P. A. C. Newman, Class. Quantum Grav. {\bf 3}, 527 (1986);
A. Ori and T. Piran, Phys. Rev. D{\bf 42}, 1068 (1990);
I. H. Dwivedi and P. S. Joshi, J. Math. Phys. {\bf 32}, 2167 (1991);
Phys. Rev. D{\bf 45}, 2147 (1992); P. S. Joshi and I. H. Dwivedi,
Gen. Rel. and Grav. {\bf 24}, 129 (1992); Commun. Math. Phys. {\bf 146},
333 (1992); Class. Quantum Grav. {\bf 9}, L69 (1992); K. Lake, Phys. Rev.
D{\bf 43}, 1416 (1991); Phys. Rev. Lett. {\bf 68}, 3129 (1992);
S. L. Shapiro and S. A. Teukolsky, Phys. Rev. Lett. {\bf 66}, 994 (1991);
Phys. Rev. D{\bf 45}, 2006 (1992).}
\item{2.}{R. Penrose, Nuovo Cimento {\bf 1}, 252 (1969).}
\item{3.}{S. Hawking, Commun. Math. Phys. {\bf 43}, 199 (1975);
Phys. Rev. D{\bf 13}, 191 (1976);
J. D. Bekenstein, Phys. Rev. D{\bf 7}, 2333 (1973); Phys. Rev. D{\bf 9}, 3292
(1974).}
\item{4.}{S. Hawking, Phys. Rev. D{\bf 14}, 2460 (1976).}
\item{5.}{For reviews, see: T. Banks, M. E. Peskin,
and L. Susskind, Nucl. Phys. {\bf B244} 125 (1984); J. Preskill, ``Do Black
Holes Destroy Information?"
California Institute of Technology preprint CALT-68-1819 (1992); S. B.
Giddings, Phys. Rev. D{\bf 46}, 1347 (1992); ``Toy Models for Black Hole
Evaporation" UCSBTH-92-36 hep-th/9209113 (1992); J. A. Harvey and A.
Strominger,
``Quantum Aspects of Black Holes", preprint EFI-92-41 hep-th/9209055 (1992).}
\item{6.}{C. Brans and R. H. Dicke, Phys. Rev. {\bf 124}, 925 (1961).}
\item{7.}{J. W. Moffat, Phys. Rev. D{\bf 35}, 3733 (1987);
For a review, see: {\it Gravitation 1990: A Banff Summer Institute, ed.
R. B. Mann and P. Wesson (Singapore: World Scientific) p. 523 1991.}
\item{8.}{D. Vincent, Class. Quantum Grav. {\bf 2}, 409 (1985).}
\item{9.}{P. Savaria, Class. Quantum Grav. {\bf 6}, 1003 (1985);
L. Campbell, J. W. Moffat, and P. Savaria, Ap. J. {\bf 372}, 241 (1991).}
\item{10.}{P. Savaria, Class. Quantum Grav. {\bf 9}, 1349 (1992).}
\item{11.}{R. Penrose, Phys. Rev. Lett. {\bf 14}, 57 (1965);
S. Hawking, Proc. R. Soc. London {\bf A300}, 187 (1967).}
\item{12.}{It has recently been  claimed that NGT suffers from radiation flux
instability
and bad asymptotic behaviour: T. Damour, S. Deser, and J. McCarthy,
Phys Rev. D{\bf 45}, R3289, (1992), preprint BRXTH-324, IHES/P/92/36,
ADP-92-180/M5 (1992). These results have been shown to be incorrect:
N. Cornish, J. W. Moffat, and D. Tatarski, Phys. Lett A {\bf 173}, 109
(1993); ``Gravitational Waves from an Axi-Symmetric Source in the Nonsymmetric
Gravitational Theory", preprint 1993 (submitted to Phys. Rev. D);
N. Cornish and J. W. Moffat, Phys. Rev. D (to be published).}
\item{13.}{J. A. Wheeler, Phys. Rev. {\bf 97}, 511 (1955);
{\it Geometrodynamics}, Academic Press Inc. New York, 1962.}
\item{14.}{J. A. Wheeler, {\it Relativity, Groups and Topology},
ed. C. and B. DeWitt (New York: Gordon and Breach 1964);
B. K. Harrison, K. S. Thorne, M. Wakano, and J. A. Wheeler,
{\it Gravitation Theory and Gravitational Collapse}, University of Chicago
Press, Chicago and London, 1965.}
\item{15.}{D. R. Brill and J. B. Hartle, Phys. Rev. {\bf 135}, B271 (1964).}
\item{16.}{D. R. Brill and J. A. Wheeler, Rev. Mod. Phys. {\bf 29}, 465
(1957).}

\end